\documentclass[12pt]{article}

\pdfoutput=1

\usepackage{graphics}
\usepackage{epsfig}

\usepackage{amsfonts}
\usepackage{amssymb}
\usepackage{amsmath}
\usepackage{dsfont}
\usepackage{multirow}
\usepackage{latexsym}
\usepackage{verbatim}
\usepackage{mcite}
\usepackage{cite}
\usepackage{units}
\usepackage[all]{xy}
\usepackage[usenames,dvipsnames,table]{xcolor}
\usepackage{cancel}
\usepackage{slashed}
\usepackage{hyperref}
\usepackage{physics}
\usepackage{breqn}

\def\beq{\begin{equation}}
\def\eeq{\end{equation}}
\def\beqa{\begin{eqnarray}}
\def\eeqa{\end{eqnarray}}
\newcommand{\f}{\frac}

\def\bfone{\relax{\rm 1\kern-.35em 1}}

\def\vr{{\mathbf{r}}}
\def\vx{{\mathbf{x}}}

\def\vp{{\mathbf{p}}}
\def\vq{{\mathbf{q}}}
\def\vk{{\mathbf{k}}}

\def\vu{{\mathbf{u}}}

\newcommand{\cB}{{\cal B}}

\newcommand{\cD}{{\cal D}}

\newcommand{\cH}{{\cal H}}

\newcommand{\cO}{{\cal O}}

\newcommand{\be}{\begin{equation}}
\newcommand{\ee}{\end{equation}}
\newcommand{\ben}{\begin{displaymath}}
\newcommand{\een}{\end{displaymath}}
\newcommand{\bea}{\begin{eqnarray}}
\newcommand{\eea}{\end{eqnarray}}

\newcommand{\bean}{\begin{eqnarray*}}
\newcommand{\eean}{\end{eqnarray*}}

\DeclareMathAlphabet{\mathpzc}{OT1}{pzc}{m}{it}

\begin{document}
\pagestyle{plain}

\begin{titlepage}
\begin{flushright}
\end{flushright}

\bigskip

    \begin{center}
		{\Large \bf{Entanglement Entropy:
\\[15pt]
Non-Gaussian States and Strong Coupling
	}}
	
	\vskip 1.5cm

	{\bf Jos\'e J. Fern\'andez-Melgarejo$^\spadesuit$\footnote{melgarejo@at@um.es}\, and Javier Molina-Vilaplana$^\blacklozenge$\footnote{javi.molina@at@upct.es}}


	\begin{center}
	    {\it $^\spadesuit$\, Universidad de Murcia, Spain}\\
	     {\it  $^\blacklozenge$\, Universidad Polit\'ecnica de Cartagena, Spain}\\	    	    
	\end{center}

	\today

    \end{center}


\vspace{2cm}

\begin{abstract}
In this work we provide a method to study the entanglement entropy for non-Gaussian states that minimize the energy functional of interacting quantum field theories at arbitrary coupling. To this end, we build a class of non-Gaussian variational trial wavefunctionals with the help of exact nonlinear canonical transformations. The calculability \emph{bonanza} shown by these variational \emph{ansatze} allows us to compute the entanglement entropy using the prescription for the ground state of free theories. In free  theories, the entanglement entropy is determined by the two-point correlation functions. For the interacting case, we show that these two-point correlators can be replaced by their nonperturbatively corrected counterparts. Upon giving some general formulae for general interacting models we calculate the entanglement entropy of half space and compact regions for the $\phi^4$ scalar field theory in 2D. Finally, we analyze the r\^ole played by higher order correlators in our results and show that strong subadditivity is satisfied. 
\end{abstract}

\end{titlepage}

\tableofcontents


\section{Introduction}\label{sec:Intro}
Quantum entanglement is a key concept that distinguishes quantum from classical physics. It amounts to a class of nonlocal correlations between subsystems that are not present in a classical system. Remarkably, a well known measure of entanglement in quantum information theory, entanglement entropy, has found numerous applications in fields such as condensed matter physics, high energy theory and gravitational physics (see \cite{Nishioka:2018khk} and references therein). Despite a huge amount of work in this direction, the characterization of entanglement entropy in strongly interacting quantum field theories (QFT's) through explicit and systematic computations, has shown to be rather intractable to do with the exception of holographic theories. Indeed, there is a limited number of field theories for which the entanglement entropy can be exactly computed. These are free field theories and strongly coupled conformal field theories (CFT's) with holographic duals. Noteworthily, for the later ones, the entanglement entropy can be computed using the Ryu–Takayanagi formula, which amounts to be one of the central entries in the AdS/CFT dictionary \cite{Ryu:2006bv, Faulkner:2013ana}.

In general terms, when an observer has only access to a subset of the complete set of observables associated with a quantum system, tracing out the degrees of freedom localized on the non-accessible region, results in a reduced density matrix that represents the knowledge about the state of the quantum system that the observer posses. Being more explicit, let us consider that, at zero temperature, the total quantum system is described by the pure ground state $\ket{\Psi}$. Then, the density matrix is given by $\rho=\ket{\Psi}\bra{\Psi}$. If one arbitrarily divides the total system into two subsystems $A$ and $B$,  the total Hilbert space is the direct product $\cH=\cH_{A}\otimes \cH_{B}$  of the two spaces corresponding to subsystems $A$ and $B$. The expectation value of an operator  $\cO_A$ acting non-trivially on $A$ is given by
\begin{equation}
\label{eq_wightman}
\expval{ \cO_A }= {\rm Tr}\,  \left[\cO_A\, \rho\right]={\rm Tr}_A\,  \left[\cO_A \, \rho_{A}\right]\,  ,
\end{equation} 
where the trace ${\rm Tr}_A$ is taken only over the degrees of freedom pertaining to the Hilbert space $\cH_A$ and the reduced density matrix $\rho_A$ is defined as
\begin{align}
\rho_A= {\rm Tr}_B\, \rho\, ,
\end{align}
by tracing out the degrees of freedom associated to the the Hilbert space $\cH_B$. Therefore, for the observer having access only to the subsystem $A$, the physical description of the total system is given by the reduced density matrix $\rho_A$. The entanglement entropy of the subsystem $A$, which measures the amount of missing information about the total system for this observer, is given by the von Neumann entropy of the reduced density matrix $\rho_A$, \emph{e.g.},
\begin{equation}
S_A= - {\rm Tr}_A\, \left[\rho_{A} \log \rho_{A}\right]\, .
\end{equation}

In QFT, the physical content of the theory is fully determined by the knowledge of its $n$-point correlation functions. Realistic operational assumptions on the observables usually impose that an observer cannot have full access to field configurations along the entire spacetime in which the QFT is defined. This implies an incomplete knowledge about correlations between points pertaining to the accessible region $A$ and those lying outside it. Entanglement entropy quantifies the amount of information on these correlations that is loss for the observer. Despite an overwhelming body of work in this context, exact results are known only for the free QFT's and those are limited. A central feature for theories in $d$ spatial dimensions, is that entanglement entropy is a quantity that depends on a short distance cut-off regulator $\epsilon$ as 
\begin{equation}
\label{eq:area_law}
S_A\sim \frac{|\partial A|}{\epsilon^{(d-1)}}\, +\, {\rm subleading}\, ,    
\end{equation}
where $|\partial A|$ amounts to the area of the boundary of region $A$. The leading divergent term is known as the area law for the entanglement entropy. The area law is due to the large number of high energy modes that induce entanglement across the boundary $\partial A$ of the accessible region. In this sense, while for any QFT there is an infinite number of degrees of freedom per unit volume, it is not possible to devise a realistic procedure to allow an observer in $A$ to resolve infinitely small distances, so a sharp distinction between inside and outside of $A$ is solved in terms of a UV-regulator. The cut-off $\epsilon$ can be viewed as a coarse graining parameter that represents to which extent the observer distinguishes the region $A$ from the rest of the system. As a result, the entanglement entropy is necessarily cutoff dependent and sensitive to the UV structure of the quantum fluctuations. In case one wishes to use the entanglement entropy to analyze UV cutoff independent properties of the theory, then one needs to explicitly establish the sub-leading corrections in \eqref{eq:area_law}. 

\medskip

In addition, for any quantum system under consideration, the most important property of the entanglement entropy is known as strong subadditivity,
\begin{equation}
S_{A} + S_{B} \geq S_{A\cup B}+S_{A\cap B}\, .
\end{equation}
This imposes that the entanglement entropy must be a concave function as the geometric parameters defining $A$ are changed. It is the strongest condition one may set for the von-Neumann entropy. For the case of two dimensional field theories, where any region $A$ amounts to a connected compact interval of length $R$ and $S_A$ can be written as a function $S(R)$, it has been shown that the converse is also true, \emph{e.g.}, the concavity of $S(R)$ implies strong subadditivity for intervals \cite{Casini:2009sr}.

It is worth to note that most of the results commented above have been explicitly computed only in the special case of free field theories. In these theories, it is widely known that their ground states are represented by Gaussian wavefunctionals and those are completely determined by the two-point functions of the theory. For the case of interacting QFT's, it is assumed that their ground states might be described in terms of non-Gaussian wavefunctionals and thus it is expected that the entanglement entropy should also depend on higher $n$-point functions.  The question is to determine which of these correlations contribute and their weight on the structure of the entanglement entropy in specific models. In this context, the differences between free and interacting QFT's theories are not evident in terms of the entanglement entropy and have not yet been elucidated completely. As a consequence, explicit results on entanglement entropy for the case of strongly interacting field theories are scarce and  much of our knowledge comes from holographic results \cite{Nishioka:2009un}. Therefore, it is natural to develop new tools to investigate how higher order correlation functions give rise to the structure of the entanglement entropy in interacting field theories.

In \cite{Hertzberg:2012mn}, the half space ground state entanglement entropy
\footnote{That is to say, the entanglement entropy of a region resulting from tracing out the half of the system. More explicitly, regions $A$ and $B$ represent half spaces and their dividing boundary amounts to a flat space of dimension $(d-1)$.} 
of the interacting $\lambda \phi^4$ and $g\phi^3$ scalar field theories at weak coupling regime, was studied using the replica trick and position space Green’s functions. The author showed that a consistent renormalization can be performed, providing finite contributions to the entanglement entropy at one loop. In \cite{Chen:2020ild},  in a quantum mechanical setting consisting on quartic perturbative perturbations on the harmonic oscillator free case, the replica trick was used to check that to first order in perturbation theory, the entanglement entropy can be computed by means of the perturbatively corrected version of the two point correlation functions of the system.

In \cite{Cotler:2015zda}, authors used a variational principle to determine a non-perturbative approximation to the half space ground state entanglement entropy of the $\lambda \phi^4$ theory at arbitrary coupling. The variational trial states used in this study were Gaussian wavefunctionals, for which the entanglement entropy can be exactly computed. Despite the Gaussian variational approximation at large values of the coupling is well defined, the approximation is only accurate up to one loop computations \cite{Hertzberg:2012mn} and large $N$ approximations in theories such as the self interacting $O(N)$ vector model. 

\medskip

To shed some light on this problem, in this work we present a non perturbative variational approach to compute the entanglement entropy of the self interacting $\lambda \phi^4$ scalar theory for arbitrary coupling. To this end, we use a class of non-Gaussian variational trial wavefunctionals non-perturbatively built through  nonlinear canonical transformations \cite{Polley:1989wf, Ritschel:1990zs,IbanezMeier:1991hm}. 

In this respect, we briefly comment on some aspects that stem from applying a variational method to QFT: generality, calculability and ultraviolet modes.

Regarding the generality of the trial state,  this must be general enough in order to capture the most salient physical features of the ground state of the theory through the variation of its parameters.  However this is not enough. That is to say, even possessing a general \emph{ansatz} for the vacuum wavefunctional of a QFT, we are interested in efficiently evaluating expectation values of operators. The calculability problem refers to our limited knowledge in evaluating non-Gaussian path integrals which forces us to restrict the set of available trial wavefunctionals to Gaussian states.  These states (that represent the exact ground state for the case of free QFTs) have provided a great amount of non-trivial results when applied to interacting field theories. Regrettably, their favorable calculability also constrains their applicability to settings in which the relevant nonperturbative physics of the system is dominated by a single condensate that changes the mass of quasiparticles. Finally, due to the interaction between the high and low momentum modes in an interacting QFT, it would be desirable for a general variational ansatz to include variational parameters that optimally integrate out the effects of high energy modes into the low energy physics.

As it will be shown, the generality and calculability \emph{bonanza} shown by the class of variational trial states used in this work, will allow us to compute the entanglement entropy using the same prescription as in the Gaussian case, which strictly depends on two-point correlators. In our case, these 2-p functions will be replaced by their nonperturbatively corrected counterparts.  Concretely, the  method of nonlinear canonical transformations shows how nonperturbative quantum corrections on the entanglement entropy of an interacting QFT can be obtained analytically and in closed form. 

The paper has the following structure. In Section \ref{sec:nlct} we introduce the NLCTs and explain the main properties of the trial non-Gaussian wavefunctionals that we build upon their application on Gaussian states. In Section \ref{sec:ee_nlct} we review the obtaining of the entanglement entropy of the half space and compact regions for Gaussian states. Additionally, we explain our method to obtain the entanglement entropy of the same subsystems for non-Gaussian states. Our results are independent of the theory. Then, in Section \ref{sec:ee-phi4} we pick the $(1+1)$-dimensional $\phi^4$ and evaluate such quantities for the trial state that minimizes the energy functional. In addition, we study the strong subadditivity condition for the case of intervals. Finally, we discuss our results and future directions in Section \ref{sec:discussion}.

\section{Non-Gaussian States through NLCT}
\label{sec:nlct}
 In the context of variational methods in QFT, Gaussian states are trial states that exactly represent the ground state of free field theories. The variational method consists of minimizing the expectation value of the Hamiltonian with respect to a set of variational parameters describing the trial wavefunctional $\Psi[\phi]$.  In particular, a Gaussian wavefunctional is parameterized by a real-valued function $\bar{\phi}(\mathbf{x})$ and a real-valued symmetric kernel $G(\mathbf{x},\mathbf{x'})$,
\begin{eqnarray}
\label{eq:gaussian_wavefunc}
	\Psi[\phi] &=& N \exp \Big[ - \frac{1}{4} \int d^d\mathbf{x} \int d^d\mathbf{y}
	\left( \phi(\mathbf{x}) - \bar{\phi}(\mathbf{x}) \right) G^{-1}(\mathbf{x},\mathbf{y}) 
	\left( \phi(\mathbf{y}) - \bar{\phi}(\mathbf{y}) \right) \Big]\\ \nonumber
	&\equiv& N\, \exp \Big[ \! - \frac{1}{4} \, ( \phi - \bar{\phi}) \, G^{-1} ( \phi - \bar{\phi}) \Big]\, ,
\end{eqnarray}
where t$d$ is the spatial dimension and $N = [\det(2\pi G)]^{-1/4}$ enforces $\Psi$ to have unit norm. In the second line we have used a more compact notation.

From the Schr\"odinger picture of field theory it follows immediately that
\begin{equation}
	\bar{\phi}(\mathbf{x}) = \langle \Psi| \, \phi(\mathbf{x}) \, | \Psi \rangle
\end{equation}
and that $G(\mathbf{x},\mathbf{y})$ is the two-point function (propagator) or Green function \cite{Cotler:2015zda}
\begin{equation}
	G(\mathbf{x},\mathbf{y}) 
	= \frac{1}{2}\, \matrixel{\Psi}{ \left(\phi(\mathbf{x}) - \bar{\phi}(\mathbf{x})\right) \, \left(\phi(\mathbf{y}) - \bar{\phi}(\mathbf{y})\right) }{\Psi}\, .
\end{equation}

Following \cite{Polley:1989wf,Ritschel:1990zs,IbanezMeier:1991hm}, extensive non-Gaussian trial wavefunctionals can be nonperturbatively built as
\begin{align}\label{eq:NG_wavefunc}
\tilde \Psi[\phi] =\langle \phi |\tilde \Psi\rangle = \langle \phi|\,  U\,| \Psi\rangle=\langle \phi |\, \exp(\cB)\, |\Psi\rangle \, ,
\end{align}
where $\Psi \equiv \Psi[\phi]$ is a normalized Gaussian state (wavefunctional) and $U\equiv\exp(\cB)$, with $\cB^{\dagger} = -\cB$ an anti-Hermitian operator that nonperturbatively adds new variational parameters to those in the Gaussian wavefunctional. Under this transformation, the expectation value of an operator $\mathcal{O}(\phi,\pi)$ w.r.t. these states amounts to the calculation of a Gaussian expectation value for the transformed operator $\widetilde{\mathcal{O}}= U^{\dagger}\, \mathcal{O}\, U$. Noteworthily, a suitable choice of $\cB$, leads  to a non-Gaussian trial state while automatically truncating the unavoidable commutator expansion arising from application of Hadamard's lemma
\begin{align}
\widetilde{\mathcal{O}}= {\rm Ad}_{\cB}\, (\mathcal{O})= e^{{\rm ad}_{\cB}}\, \mathcal{O}\, .
\label{eq:conm_series}
\end{align}
As a result, the calculation of any expectation value with $\widetilde{\Psi}$ reduces to the computation of a finite number of Gaussian expectation values. In addition, the exponential nature of $U$ ensures an extensive volume dependence of observables such as the  energy of the system. Moreover, as $U$ is unitary, the normalization of the state is preserved. 

The operator $\cB$ consists of a product of $\pi$'s and $\phi$'s, given by
\begin{align}\label{eq:BBos}
\mathcal{B}
= 
-s\int_{\vp \vq_1\cdots \vq_m} 
h(\vp,\vq_1,\ldots,\vq_m)\, \pi(\vp)\, \phi(\vq_1)\ldots\phi(\vq_m) \delta(\vp+\vq_1+\cdots\vq_m) 
\ ,
\end{align}
with $m \in \mathbb{N}$, $\vp,\, \vq_i$ $d$-dimensional momenta and  $\int_{\vp}=\int\, d^d\vp/(2\pi)^d$. We denote these operators symbolically as $\mathcal{B}\equiv \pi\, \phi^{m}$. Here, $s$ is a  variational parameter that tracks the deviation of any observable from the Gaussian case. In order to ensure an efficient truncation of the operator expansion given by the nested commutator series appearing in Hadamard's lemma, the variational function $h(\vp,\vq_1,\ldots,\vq_m)$  is introduced in the ansatz and it must be optimized upon energy minimization. This function is symmetric w.r.t. exchange of $\vq_i$'s  and truncation imposes it has to satisfy:
\begin{align}
\label{eq:constraint}
h(\vp,\vq_1,\ldots,\vq_m)
=0 
\ , 
\quad \vp=\vq_i\, ,
\qquad\text{and} \qquad
h(\vp,\vq_1,\ldots,\vq_m)\, h(\vq_i,\vk_1,\ldots,\vk_m) 
=&\ 0
\ .
\end{align}

A suitable way of accomplishing \eqref{eq:constraint} is taking  
\begin{align}
h(\vp,\vq_1,\ldots,\vq_m)\ = g(p,q_1,\ldots,q_m)\, \eta(\vp)\cdot \kappa(\vq_1) \cdot  \kappa(\vq_2)\cdots  \kappa(\vq_m)\, ,
\label{eq:h-variational}
\end{align}
where $g(\vp,\vq_1,\ldots,\vq_m)$ is scalar function to be determined by the energy minimization and we have imposed that $\eta(\vp)\cdot \kappa(\vp) = 0$, \emph{i.e.}, the domains of momenta where $\eta$ and $\kappa$ are different from zero must be disjoint, up to sets of measure zero. Authors in \cite{Polley:1989wf, Ritschel:1990zs} provided the useful ansatz for $\eta$ and $\kappa$  given by 
\begin{eqnarray}
\label{eq:cutoff_ansatz}
\eta(\vp)&=&\Gamma((\vp/\Delta_0)^2)\, , \\ \nonumber
\kappa(\vq_i) &=& \left[\Gamma((\Delta_0/\vq_i)^2)-\Gamma((\Delta_1/\vq_i)^2)\right]\, ,
\end{eqnarray}
where  $\Delta_{0}$ and $\Delta_1$ are variationally optimized, coupling dependent momentum cutoffs and  $\Gamma(x)\equiv \theta(1-|x|)$ with $\theta(x)$ the Heaviside step function. Generically, one might understand $h(\vp,\vq_1,\ldots,\vq_m)$
as separating the Fourier components of the field $\phi$ into two non-overlapping domains of $\kappa$-``high'' and $\eta$-``low'' momenta. The coupling dependent momentum cutoffs variationally determine the size of these non-overlapping regions in momentum space in order to improve the estimation of the ground state energy. This fact is relevant in strongly-coupled theories in which the Gaussian quasi-particle picture is no longer valid.

The above constraints ensure that the commutator series terminates after the first nontrivial term. Namely, the action of $U$ on the canonical field operators $\phi(\vp)$ and $\pi(\vp)$ is given by
\begin{eqnarray}
\label{trasfields}
	\tilde \phi(\vp)&\equiv& U^\dagger\, \phi(\vp)\, U 
	=
	\phi(\vp)
	+s \, \bar{\phi}(\vp)
	\ ,
	\\ \nonumber
	\tilde \pi(\vp)&\equiv&U^\dagger\, \pi(\vp)\, U
	=
\pi(\vp)
	-\, s \, \bar{\pi}(\vp)
	\ ,
\end{eqnarray}
where the quantities with a bar are defined as the nonlinear field functionals, 
\begin{equation}
\begin{aligned}
\bar{\phi}(\vp)\equiv&\ \int_{\vq_1\cdots \vq_m} h(\vp,\vq_1,\ldots,\vq_m)\, \phi(\vq_1)\cdots \phi(\vq_m) \delta(\vp-\vq_1-\cdots\vq_m) \ , 
\\ 
\bar{\pi}(\vp)
	\equiv&\
	m\, \int_{\vq_1\cdots \vq_m}
	h(\vq_1,\vp,\ldots,\vq_m)\, \pi(\vq_1)\, \phi(\vq_2)\phi(\vq_m) 
	\delta(\vp-\vq_1-\cdots\vq_m) 
\ .
\end{aligned}
\label{detail_transfields}
\end{equation}

A central consequence of $U$ being unitary is that the canonical commutation relations (CCR) still hold under the nonlinear transformed fields \eqref{trasfields} and \eqref{detail_transfields} giving,
\begin{align}
 [\tilde \phi(\vp), \tilde \pi(\vq)]=i\bar{\delta}(\vp + \vq)\, .
\end{align}
For this reason, the  transformations above are known as nonlinear canonical transformations (NLCT). As commented above, the consequence of Eq. \eqref{trasfields} is that the expectation value of an arbitrary operator $\mathcal{O}(\pi, \phi)$ w.r.t. $\widetilde{\Psi}$, reduces to a Gaussian expectation value for the transformed operator $\widetilde{\mathcal{O}}=U^{\dagger}\, \mathcal{O}\, U$, as we have
\begin{align}
\langle \tilde \Psi |\mathcal{O} | \tilde \Psi\rangle = \langle \Psi |U^{\dagger}\, \mathcal{O}\, U | \Psi\rangle \equiv \expval{U^{\dagger}\, \mathcal{O}\, U }\, .
\end{align} 
Throughout this paper we will consider two types of expectation values, which we will denote
\begin{align}
\expval{\cO}
\equiv
\langle  \Psi |\mathcal{O} | \Psi\rangle
\ ,
\qquad\qquad
\expval{\cO}_{\tilde\Psi}
\equiv
\langle \tilde \Psi |\mathcal{O} | \tilde \Psi\rangle
\ ,
\end{align}
 
When considering QFT, $n$-point correlation functions of the form  $\expval{\phi(\vp_1)\cdots\phi(\vp_n)}_{\widetilde{\Psi}}$ are of particular interest. Using \eqref{trasfields} and \eqref{detail_transfields}, one obtains

\begin{dmath}[spread={5pt}]
\expval{\phi(\vp_1)\cdots\phi(\vp_n)}_{\widetilde{\Psi}}
	=
	\expval{U^\dagger\, \phi(\vp_1)\, U \cdots U^\dagger\, \phi(\vp_n)\, U}
	=\ \expval{\phi(\vp_1)\cdots \phi(\vp_n)}\\ \nonumber
	+ s \left[\expval{\bar{\phi}(\vp_1)\phi(\vp_2)\cdots \phi(\vp_n)} +\cdots +\expval{\phi(\vp_1)\cdots \phi(\vp_{n-1})\bar{\phi}(\vp_n)}	\right]\\
	+s^2\left[ \expval{\bar{\phi}(\vp_1)\bar{\phi}(\vp_2)\phi(\vp_3)\cdots \phi(\vp_n)}+\cdots +\expval{\phi(\vp_1)\cdots\bar{\phi}(\vp_{n-1})\bar{\phi}(\vp_{n})}\right]
\\
\mbox{} \ \  \, \, \vdots
\\
	+ s^n \expval{\bar{\phi}(\vp_1)\cdots\bar{\phi}(\vp_n)}
	\ .
\end{dmath}
This is a remarkable property of NLCT of the form $\pi\, \phi^m$, as non-Gaussian corrections to Gaussian correlation functions can be obtained in terms of a finite number of Gaussian expectation values. In particular, the terms proportional to $s^j$ in the non-Gaussian $n$-point correlation function correspond to $(n+m(j-1))$-point  Gaussian correlators, where $j=0,\ldots,n$.

Finally, we analyze the effect of the transformation $U=\exp (\mathcal{B})$ on wavefunctionals representing the probability amplitude for concrete field configurations. When applied to an interacting field theory such as the $\lambda \phi^4$ theory, the half-mean width of a variational Gaussian functional such as,
\begin{align}
\Psi[\phi]
=
N \exp\left(
-\f{1}{4} \int_\vk  \phi(\vk)\,  G^{-1}(\vk)\, \phi(-\vk)
\right)
\ ,
\end{align}
 is $(\vk^2+\mu^2)^{-1/4}$, where $\mu$ is the variational mass given by the gap equation, 
\begin{align}
\label{eq:gap_eq}
\mu^2 = m^2 + \frac{\lambda}{2}\left(I_0(\mu^2) + \bar{\phi}^2 \right)\, ,
\end{align}
where $I_N(\mu^2)=\frac12\int_\vk (\vk^2+\mu^2)^{N-\frac12}$ and $m$ and $\lambda$ are the bare mass and the bare coupling respectively. As the variational mass $\mu$ increases with interactions, then the nonclassical configurations accounted by the wavefunctional are much strongly suppressed than in the free case \cite{Ritschel:1990zs}.

To illustrate the action of ${U}$ on these Gaussian wavefunctionals, we choose the transformation $\mathcal{B}=\pi\, \phi^2$ for clarity. Noting that \cite{Ritschel:1990zs}\footnote{This can be seen by writing $U$ in the functional Schr\"{o}dinger picture, \emph{i.e.}, $\pi(\vk)\to -i\delta/\delta \phi(-\vk)$.}
\begin{align}
\Phi(\vk)\equiv {U}\, \phi(\vk) = \phi(\vk) -s\, \bar{\phi}(\vk)\, ,
\label{eq:nl_pmeasure}
\end{align}
where $\bar{\phi}(\vk)$ corresponds to \eqref{detail_transfields} for $m=2$, we obtain
\begin{equation}
\begin{aligned}
	\tilde \Psi[\phi] 
\equiv&\ 
U\, \Psi[\phi] = U\, \left(1 - \f{1}{4} \int_\vk \phi(\vk) G^{-1}(\vk) \phi(-\vk) +\cdots\right)
\\
=&
\ 1-	\f{1}{4} \int_\vk \left(	\phi(\vk)	-s \bar{\phi}(\vk)\right)
 G^{-1}(\vk)\left(	\phi(-\vk)	-s\bar{\phi}(-\vk)\right)+\cdots
\\ 
=&\
1-\f{1}{4} \int_\vk \Phi(\vk)\,  G^{-1}(\vk)\, \Phi(-\vk) +\cdots 
\\ 
=&\
\Psi\left[\Phi\right]
	\ ,
\end{aligned}
\label{eq:effective-gaussian}
\end{equation}
where ellipses stand for the expansion of the exponential. 

As a result, the non-Gaussian trial state $\tilde \Psi[\phi]$ can be understood as an \emph{effective} Gaussian state $\Psi[\Phi]$ on a set of fields $(\Phi)$ that are a nonlinear deformation \eqref{trasfields} of the elementary \emph{microscopic} fields  $(\phi)$ that appear in the Hamiltonian which defines the theory under consideration. In other words, $U$ generates a translation of the argument in the configuration space of the theory that symbolically reads as $\tilde \Psi[\phi]=\Psi[\phi-s \phi^m]$ for arbitrary $m$. Thus, we are considering a class of field transformations that shift part of the degrees of freedom of $\phi$ by a nonlinear polynomial function of other degrees of freedom which remain unaffected by the transformation. 

A crucial point for forthcoming discussions in this work refers to the Jacobian of these NLCT and its influence in the path integral measure, given by  $\cD\Phi = {\rm det} J\, \cD\phi$. Using ${\rm det}\, J= e^{{\rm Tr}\log\, J}$, for the class of nonlinear transformation in Eq.\eqref{eq:nl_pmeasure}, the result is $J=1 + M$, where, under very general circumstances\footnote{Those refer to assuming that $|s|\ll 1$. As it will be shown later this does not does not necessarily imply that we are considering $\lambda\ll1$}, $M$  allows for a convergent expansion  of $\log(1+M)= M -M^2/2! + M^3/3! \cdots$. Remarkably, Eq. \eqref{eq:constraint} imposes both $M^k=0$ for $k>1$ and ${\rm Tr}\, M = 0$ and thus ${\rm det}\, J=1$. This leaves the path integral measure invariant $\cD\Phi = \cD\phi$ (see \cite{IbanezMeier:1992nc} for complementary comments on the invariance of the path integral measure due to the class of NLCT considered here).

To summarize, nonlinear canonical transformations yield variational non-Gaussian trial wavefunctionals by applying the operator $U=\exp\left(\mathcal{B}\right)$ defined through the variational function $h(\vp,\vq_1,\ldots,\vq_m)$ to a Gaussian wavefunctional with a variational kernel $G(\vp)$. As this is a model independent formalism, the explicit dependence of the variational parameters on the  couplings of a theory has to be established through energy minimization. This will be illustrated in Section \ref{sec:ee-phi4} for a concrete example.

\section{Entanglement Entropy of Non-Gaussian States}
\label{sec:ee_nlct}
In general, given a wavefunctional $\Psi[\phi(\vx)]$ of the QFT under consideration, the entanglement entropy of a region $A$ is computed through the reduced density matrix $\rho_A[\Psi] = \Tr_B \left[ \Psi[\phi] \Psi^*[\phi]\right]$ by tracing out the degrees of freedom in the spacetime region $B$ complementary to $A$.

In this section we are going to review the entanglement entropy of half space and compact arbitrary regions in free theories. Being only dependent on 2-point correlators, we are going to make use of the property \eqref{eq:effective-gaussian} to calculate and analyze the entanglement entropy of non-Gaussian states that minimize the energy functional of generic interacting theories.

\subsection{Entanglement Entropy of Gaussian States}\label{ssec:ee_gaussian}
For the sake of clarity in exposition, we describe here the two most general methods to calculate the entanglement entropy of Gaussian states, the replica trick and the real-time approach (see \cite{Nishioka:2018khk, Casini:2009sr} for excellent reviews). 
 
 \subsubsection*{Half space}
When the region of interest amounts to the half-space, the entanglement entropy of Gaussian states can be easily computed through the replica trick \cite{Callan:1994py}
\begin{equation}
\label{replicaa}
S\left[\rho_A\right] \equiv \lim_{n \to 1}\left( - \frac{d}{dn} + 1 \right) \log(\Tr \left[\rho_A\right]^n) = - \Tr \left(\rho_A \log \rho_A\right)\, .
\end{equation} 

As said, being $A$ the half-space, it can be shown that
\begin{align}
\Tr \left[\rho_A\right]^n \propto Z_\delta\, ,
\end{align}
with $Z_\delta$ being the partition function of a massive free scalar field of mass $m$ living on an $n$-sheeted conical Riemann surface with a deficit angle $\delta = 2 \pi (1-n)$.  When $A$ is the half plane, $\frac{d}{dn} = - 2 \pi \frac{d}{d\delta}$ and \eqref{replicaa} reads
\begin{equation}
\label{replicab}
S\left[\rho_A\right]  \equiv\lim_{\delta \to 0} \left(2 \pi \frac{d}{d\delta} + 1\right) \log Z_{\delta}\, .
\end{equation}
The result for the class of the Gaussian wavefunctionals under consideration can be symbolically written as \cite{Callan:1994py, Solodukhin:2011gn}
\begin{align}
S\left[\rho_A\right] = -\frac{|\partial A|}{12}\, \log\, \det\, [\vp^2 +m^2] + {\rm const}\, ,
\label{eq:ee_gaussian_rt} 
\end{align}
where the constant accounts for the normalization of the Gaussian wavefunctional \eqref{eq:gaussian_wavefunc}. Noting that 
\begin{align}
\label{eq:gaussian_kernel}
G(\vp) = \frac{1}{2\, \sqrt{\vp^2 + m^2}}\, ,
\end{align}
one may write
\begin{dmath}
S\left[\rho_A\right]  
=
-\frac{|\partial A|}{12}\, \log\, \det\, \frac{1}{4}\, G(\vp)^{-2} + {\rm const}
= 
\frac{|\partial A|}{6}\, \int_{\vp}\, \log\, \expval{\phi(\vp)\, \phi(-\vp)} + {\rm const} 
\, .
\label{eq:ee_gaussian_half} 
\end{dmath}

\subsubsection*{Compact arbitrary regions}

We are going to study the entanglement entropy of a compact region $A$ for Gaussian states. While this can done using the replica trick as well \cite{Holzhey:1994we,Calabrese:2004eu,Calabrese:2009qy}, here we will use the real time approach which was the first method used to compute entanglement entropy in free theories \cite{Bombelli:1986rw}. In this method, which has been mainly applied to numerical calculations in the lattice, the idea is to obtain the reduced density matrix $\rho_A$ in terms of the two point correlators restricted to the region $A$.  As a quantum field theory can be completely specified in terms of its correlation functions, Eq. \eqref{eq_wightman} implies that the information encoded by all the correlators inside $A$ fully determines the density matrix $\rho_A$. For the case of Gaussian states this is very simple, as the Wick theorem imposes that the only nontrivial correlators are the two point correlation functions. This was exploited in \cite{Peschel2003,Araki:1971id} to derive explicit expressions of $\rho_A$ in free boson and fermion discrete systems in terms of correlators. 

Let us briefly review this approach for the case of (discretized) free boson  theories \cite{Casini:2009sr}. We denote the local field variables $\phi _{m}$ and $\pi _{n}$  with canonical commutation relations 
\begin{equation}
[\phi _{m},\pi _{n}]=i\delta _{mn}\,,\hspace{1cm} [\phi _{m},\phi _{n}]=[\pi _{m},\pi _{n}]=0\,.
\end{equation} 
The two point correlators of the field variables inside $A$ are given by
\begin{equation}
\begin{aligned}
\left\langle \phi _{m}\phi _{n}\right\rangle \equiv &\ X_{mn} \,, \hspace{1cm}\left\langle \pi _{m}\pi _{n}\right\rangle \equiv P_{mn} \,, \\ 
 \left\langle \phi _{m}\pi _{n}\right\rangle =& \left\langle \pi _{n}\phi _{m}\right\rangle^*=\frac{i}{2}\delta _{mn}\, ,
\end{aligned}
\end{equation}
where $X_{mn}$ and $P_{mn}$ are real Hermitian and positive matrices. It can be shown that the entanglement entropy in region $A$ is given in terms of the (positive) eigenvalues of $C=\sqrt{XP}$ as
\begin{equation}
S\left[\rho_A\right]  =\Tr\Big[(C+1/2)\log (C+1/2)-(C-1/2)\log (C-1/2)\Big]\, . \label{eq:entr_rt}
\end{equation}

For free quadratic bosonic Hamiltonians given by    
\begin{align}
H
=
\frac{1}{2}\sum_m \pi _{m}^{2}+\frac{1}{2}\sum_{mn}\phi _{m}\Omega_{mn}\phi_{n}
\ ,
\end{align}
the ground state correlators are  
\begin{eqnarray}
X_{mn} &=&\frac{1}{2}(\Omega^{-\frac{1}{2}})_{mn}\,,    \nonumber\\
P_{mn} &=&\frac{1}{2}(\Omega^{\frac{1}{2}})_{mn}\,. 
\label{corr_xp}
\end{eqnarray}

The matrix $C =\sqrt{XP}$ is constant, $C = 1/2$, when region $A$ amounts to the total system, and thus, the entanglement entropy in Eq. \ref{eq:entr_rt}
vanishes. In general, it does not vanish for an arbitrary compact region $A$ specified by $i = 1, \cdots\, R$ where  $R< N$ ($N$, the total number of sites in the system) as far as
\begin{align}
C_{ij}=\frac{1}{2}\sqrt{\sum_{k=1}^R\, (\Omega^{-1/2})_{ik}\, (\Omega^{1/2})_{kj}}\, ,
\end{align}
is not necessarily $1/2$ in general.

To give a simple example \footnote{Again, we emphasize that this prescription can be trivially generalized to any dimension.}, let us specify the Hamiltonian and correlators that must be used in a one dimensional lattice to calculate numerically the entanglement entropy for a real massive scalar:
\begin{equation}
H=\frac{a}{2}\sum_{n=-\infty}^{\infty}\left( \pi _{n}^{2}+(\phi _{n+1}-\phi
_{n})^{2}+\,m^2\phi _{n}^{2}\right) \,,
\end{equation}
where $a$ is the lattice spacing. In this setting, the Fourier modes of the field $\phi(p)$ are periodic functions of momentum restricted to the first Brillouin zone $-\pi/a \leq p \leq \pi/a$ with the propagator \cite{ZinnJustin:2002ru} given by
\begin{align}
G^{-1}(p)=\left(m^2 + \frac{2}{a^2}\left[1- \cos (a p) \right]\right)^{1/2}\, .
\end{align} 
With this, the correlators \eqref{corr_xp} can be written as
\begin{equation}
\begin{aligned}
\expval{\phi_{n}\phi_{m}} &=& \int_{-\pi/a}^{\pi/a}\, \frac{dp}{2\pi}\, e^{ i \, p\, a\, (m-n)}\, \frac{1}{2\, G(p)}\, 
\\ 
\expval{\pi_{n}\pi_{m}} &=& \int_{-\pi/a}^{\pi/a} \, \frac{dp}{2\pi}\, e^{ i\, p\, a\, (m-n)}\, \frac{G(p)}{2}\,.
\end{aligned}
\end{equation}
Further details of this method can be found in \cite{Casini:2003ix}.

\subsection{Entanglement Entropy of Non-Gaussian States through NLCT}
\label{ssec:ee_nlct}

Having discussed the Gaussian case, now we focus in NLCT wavefunctionals which, as discussed above, take the \emph{effective} Gaussian form 
\begin{align}
\widetilde{\Psi}[\phi] = \Psi[\Phi]\, .
\end{align}
We note that the reduced density matrix is Gaussian w.r.t. the nonlinear canonically transformed fields, i.e.,
\begin{dmath}
\rho_A[\tilde \Psi]
=
\Tr_B\Big[ \tilde \Psi[\phi]\, \tilde \Psi^{*}[\phi]\Big]
= \int \mathcal{D}\phi_B\, \widetilde{\Psi}[\phi_A,\phi_B]\, \widetilde{\Psi}^*[\phi'_A,\phi_B]
= \int \mathcal{D}\Phi_B\, \, \Psi[\Phi_A,\Phi_B]\, \Psi^*[\Phi'_A,\Phi_B] = 
\Tr_B\Big[ \Psi[\Phi]\, \Psi^{*}[\Phi]\Big]\, ,
\end{dmath} 
where the third equality is due to the invariance of the path integral measure $\mathcal{D}\phi = \mathcal{D}\Phi$ under NLCT commented above. 

With this, in order to compute the entanglement entropy of a NLCT non-Gaussian state we apply, and this is the central hypothesis of this work, the techniques for Gaussian states being discussed above. That is to say, as for Gaussian states, the entanglement entropy is fully determined through the two-point correlation functions, our proposal is to use the computational structure of the Gaussian free case but with the two-point correlators replaced by the nonperturbatively-corrected counterparts yielded by a nonlinear canonical transformation.

\subsubsection*{Half space}
 
Regarding the entanglement entropy of the half-space for a Gaussian wavefunctional, Eq. \eqref{eq:ee_gaussian_half}, a precise statement on the hypothesis presented above is that the entanglement entropy of non-Gaussian wavefunctional reads
\begin{dmath}
 S[\rho_A[\tilde \Psi]] 
 = \frac{|\partial A|}{6}\, \int_{\vp}\, \log \, \expval{\phi(\vp)\, \phi(-\vp)}_{\tilde \Psi}+ {\rm const}
 = \frac{|\partial A|}{6}\, \int_{\vp}\, \log \, \expval{\tilde \phi(\vp)\, \tilde \phi(-\vp)}+ {\rm const} \, ,
\label{eq:half-space-NG}
\end{dmath}
where
\begin{dmath}
 \expval{\tilde \phi(\vp)\, \tilde \phi(-\vp)} 
 = \expval{\phi(\vp)\, \phi(-\vp)} 
  + s^2 \expval{\bar{\phi}(\vp)\, \bar{\phi}(-\vp)}\, . 
\end{dmath}

Then, plugging these correlators into the expression for $S[\rho_A[\tilde \Psi]]$ \eqref{eq:half-space-NG}, we have
\begin{dmath}
S[\rho_A[\tilde \Psi]]/|\partial A|
=
\frac{1}{6}\int_\vp \log\expval{\phi(\vp)\phi(-\vp)}
+\frac{s^2}{6}\int_\vp \frac{\expval{\bar\phi(\vp)\bar\phi(-\vp)}_{m}}{\expval{\phi(\vp)\phi(-\vp)}}
+\cO(s^4)
\ ,
\label{eq:ee-halfspace-expansion}
\end{dmath}
where we have assumed that $|s|<1$. As it will be made clear later, this assumption encompasses regimes of the physical coupling ranging from small to very large values.

For the case $m=2$, the transformation in momentum space is given by \cite{Fernandez-Melgarejo:2019sjo,Fernandez-Melgarejo:2020fzw}
\begin{dmath}
\expval{\bar\phi(\vp)\bar\phi(-\vp)}_{2}
=
\frac{1}{2\eta^2}
	\int_{\vq_{1,2}} h \left(|\vq_1+\vq_2|,\vq_1,\vq_2 \right)^2 \, G(\vq_1)\, G(\vq_2)\, \delta(\vp-\vq_1-\vq_2)
\ .
\label{eq:2point-NG}
\end{dmath}
Here, the $h$ function is given in \eqref{eq:h-variational}.

Again, we emphasize that for an state $\tilde\Psi$, this expression is valid for any $d$-dimensional interacting theory. Such integrals must be evaluated when considering the optimal parameters that minimize the energy functional, which are precisely the  objects that carry the information about one particular theory (see \cite{Ritschel:1990zs}).

\subsubsection*{Compact arbitrary regions}

In terms of the real time approach, the entanglement entropy of Gaussian states for an area $A$ is given by \eqref{eq:entr_rt}.  As commented above, for a NLCT of the form $\pi\, \phi^m$, non-Gaussian corrections to Gaussian correlation functions can be obtained in terms of a finite number of Gaussian expectation values. In particular, as we focus on 2-point functions, the terms proportional to $s^j$ in the non-Gaussian $2$-point function correspond to $(2+m(j-1))$-point Gaussian correlators, where $j=0,1,2$. In this sense, that is how our method includes the effects of higher order correlations in the computation of entanglement entropy.

Then, based on the property \eqref{eq:effective-gaussian} for $\tilde\Psi[\phi]$, we can apply the entanglement entropy expression \eqref{eq:entr_rt} for the correlators associated to the non-Gaussian states. That is to say, our proposal amounts to 
\begin{equation}
S[\rho_A[\tilde \Psi]] 
=
\Tr\Big[(\tilde C+1/2)\log (\tilde C+1/2)-(\tilde C-1/2)\log (\tilde C-1/2)\Big]\, , \label{eq:entr_rt_nlct}
\end{equation} 
where the tilded correlators are
\begin{eqnarray}
\left\langle \tilde \phi _{i} \tilde \phi _{j}\right\rangle &\equiv& \tilde X_{ij} \,, \hspace{1cm}\left\langle \tilde \pi _{i} \tilde \pi _{j}\right\rangle \equiv \tilde P_{ij} \,, \\ \nonumber
 \left\langle \tilde \phi _{i} \tilde \pi _{j}\right\rangle &=& \left\langle \tilde \pi _{j} \tilde \phi _{i}\right\rangle^*=\frac{i}{2}\delta _{ij}\, ,
\end{eqnarray}
and $\tilde C \equiv \sqrt{\tilde X \tilde P}$.

These non-Gaussian 2-point correlators are obtained from the nonlinear transformation \eqref{detail_transfields}. For $m=2$, they are given by
\begin{align}
\tilde X_{ij}
=&\
X_{ij}
+
\frac{s^2}{2\eta^2} \int_{\vp\vq} e^{i(\vp\cdot \vu_i+\vq\cdot\vu_j)a}
h(|\vp+\vq|,\vp,\vq)^2\, G(\vp)\, G(\vq)
\ ,
\\
\tilde P_{ij}
=&\
P_{ij}
+
\frac{s^2}{\eta^2}\int_{\vp\vq} e^{i(\vp\cdot\vu_i+\vq\cdot\vu_j)a}
h(|\vp+\vq|,\vp,\vq)^2\, \frac{G(\vq)}{G(\vp)}
\ ,
\end{align}
where $\vu_i$ is a unitary vector that locates the site of the lattice and $X_{ij}$ and $P_{ij}$ written in \eqref{corr_xp}.

As in the half space case, we have obtained these expressions in full generality without specifying what is the interacting theory nor any dimensionality.

In the following section we are going to apply these formulas to a particular theory, which translates to picking a state that minimizes the energy through a variational method.

\section{Entanglement Entropy in $\phi^4$ Theory}
\label{sec:ee-phi4}

In this section we calculate the entanglement entropy of non-Gaussian states generated by NLCT that minimize the energy functional of the $(1+1)$-dimensional $\lambda\phi^4$ theory, which has the following Hamiltonian density:
\begin{align}
\label{eq:hamiltonian}
\cH= \frac{1}{2} \pi^2 + \frac{1}{2}(\nabla \phi)^2 + \frac{m^2}{2} \, \phi^2 + \frac{\lambda}{4!} \, \phi^4 \, .
\end{align}

In $d = 1$, this theory does not exhibit any issue when renormalization is considered, and thus the method and our results lie on a solid ground. For higher dimensions, however, things are more subtle. Additional divergences arise from the non-Gaussian terms and they should be properly addressed \cite{Polley:1989wf,Ritschel:1990zs}.

In $d=1$, it is known that this model experiences a second order phase transition at which the vacuum changes continuously from a symmetric to a nonsymmetric state \cite{Chang:1976ek}. This transition cannot be detected by perturbation theory and occurs at strong coupling. The rigorous proof of this fact \cite{McBryan:1976ga} does not allow to compute the critical coupling. An estimate was obtained by the variational Gaussian approximation \cite{Chang:1975dt}, but this yields a wrong critical behavior as it predicts a first order phase transition. The variational method based on the non-linear canonical transformations (NLCT) was used to detect this second order phase transition while computing the critical value of the coupling constant \cite{Polley:1989wf,Ritschel:1990zs}. 

In the perturbative regime, for the self interacting scalar theory, given a NLCT of the form $\pi\, \phi^n$, it is possible to systematically obtain a diagrammatic interpretation of the contributions to the $n$-point correlation functions yielded by the ansatz. This gives account of the non-Gaussian contributions to the correlators  \cite{Ritschel:1990zs, Fernandez-Melgarejo:2020fzw}. As it is shown in these references, the diagrammatic content of the correlation functions yielded by NLCT trial states adds up a much larger class of Feynman diagrams than the usual "cactus"-like ones of the Gaussian approach.

Here we circumscribe to a treatment of the the theory in terms of a $\pi\, \phi^2$ transformation. The equations for the optimal values of the variational parameters  $s$, $G(\vp)$ and $h(\vp,\vq_1,\vq_2)$ are
obtained, for a fixed $\phi_c = \expval{\phi(x)}_{\tilde \Psi}$, by deriving $\expval{\cH}_{\tilde \Psi}$ w.r.t. them and then equating to zero (details can be found in \cite{Polley:1989wf, Ritschel:1990zs}). This yields a set of nonlinearly coupled equations that  greatly simplify for $\phi_c \sim 0$. 

Indeed, in this case, the kernel $G(\vp)$ reduces to the Gaussian case \eqref{eq:gaussian_kernel} up to order $\mathcal{O}(\phi_c^2)$, with a variational mass $\mu$ given by the gap equation \eqref{eq:gap_eq}, the optimal $s = -4\, \lambda\, \phi_c$ and
\begin{align}
g(|\vq_1+\vq_2|,\vq_1,\vq_2)
=
\frac{4}{G^{-1}(\vq_1+\vq_2)\left[ G^{-1}(\vq_1) + G^{-1}(\vq_2)\right]}\, .
\label{eq:var_g}
\end{align}
Finally, with the results, numerical optimization is used in order to find the optimal values of the coupling-dependent cutoffs $\Delta_0$ and $\Delta_1$.

Here, we carry out an optimization in which we have fixed $\phi_c=10^{-2}$. This implies that our formulas are valid for $\lambda\le100$ in such a way we keep $|s|<1$.

\subsubsection*{Half space}
\label{ssec:half-space}

\begin{figure}[!t]
\centering
\includegraphics[width=.8\textwidth]{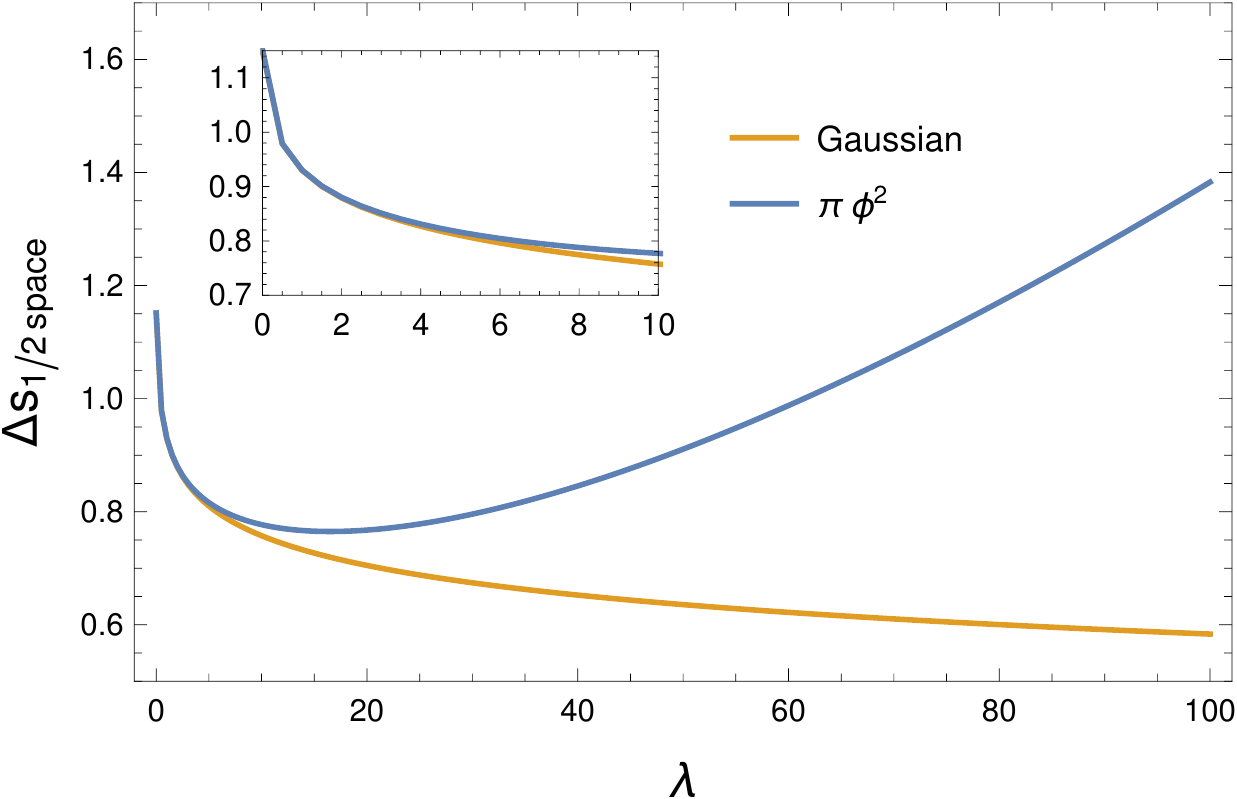}
\caption{\textit{
Finite contributions to the half-space entanglement entropy per unit area $\Delta s$ as a function of the coupling parameter $\lambda$ for the Gaussian state $\Psi[\phi]$ (orange) and the non-Gaussian states $\tilde\Psi[\phi]$ generated by the NLCT with $m=2$ (blue).
}}
\label{fig:half-space}
\end{figure}

For the non-Gaussian transformation \eqref{eq:BBos} with  $m=2$, the energy minimization requires the optimal values of the variational parameters to be $\Delta_0=0.031$ and  $\Delta_1=0.97$. When fixing such values, we numerically evaluate the finite part of $S[\rho_A[\tilde \Psi]]/|\partial A|$. 

In Figure \ref{fig:half-space} we plot $\Delta s_{NG}$, with
\begin{align}
\Delta s_{NG}
\equiv
\frac{S[\rho_A[\tilde \Psi]]- S_{G,div}}{|\partial A|}
\ ,
\end{align}
where $S_{G,div}$ denotes the divergent terms of the Gaussian contribution, namely the divergences of the first term in \eqref{eq:ee-halfspace-expansion} (see \cite{Hertzberg:2010uv, Hertzberg:2012mn}). That is to say, we plot the finite part of the entanglement entropy per unit area of non-Gaussian states as a function of the coupling $\lambda$ for states generated by the  $m=2$ NLCT. For comparison, we also plot this quantity for the Gaussian state. 

Some comments are in order. In agreement with \cite{Cotler:2015zda}, $\Delta s_{NG}$ for the Gaussian state decreases as the coupling grows. Regarding the non-Gaussian case, at weak coupling the entanglement entropy shows a similar profile than the Gaussian state. On the other hand, it notably diverts from the Gaussian entropy for values $\lambda>1$. At this regime, this seems to indicate that correlators beyond 2p-functions become relevant, as the entanglement entropy for Gaussian states only captures the features of the 2-point connected correlators.

For a more detailed discussion of our results, we refer to Section \ref{sec:discussion}.

\subsubsection*{Intervals}
\label{ssec:intervals}
 Connected compact regions in $d=1$ are intervals determined, up to a translation, by their length
$R$. For these intervals the entropy can be written as a function of a real variable $S(R)$.
We now calculate $S(R)$ for non-Gaussian wavefunctionals generated by the NLCT transformations \eqref{eq:BBos}.

Upon picking the same optimization scheme as in the previous section (we will consider $m=2$), we find the variational parameters $\Delta_0$ and $\Delta_1$ that minimize the energy functional for a fixed $\lambda$. Then we use these values to build the non-Gaussian state and thus evaluate the correlators in the entropy formula \eqref{eq:entr_rt_nlct}.

\begin{figure}[t]
\centering
\includegraphics[width=.8\textwidth]{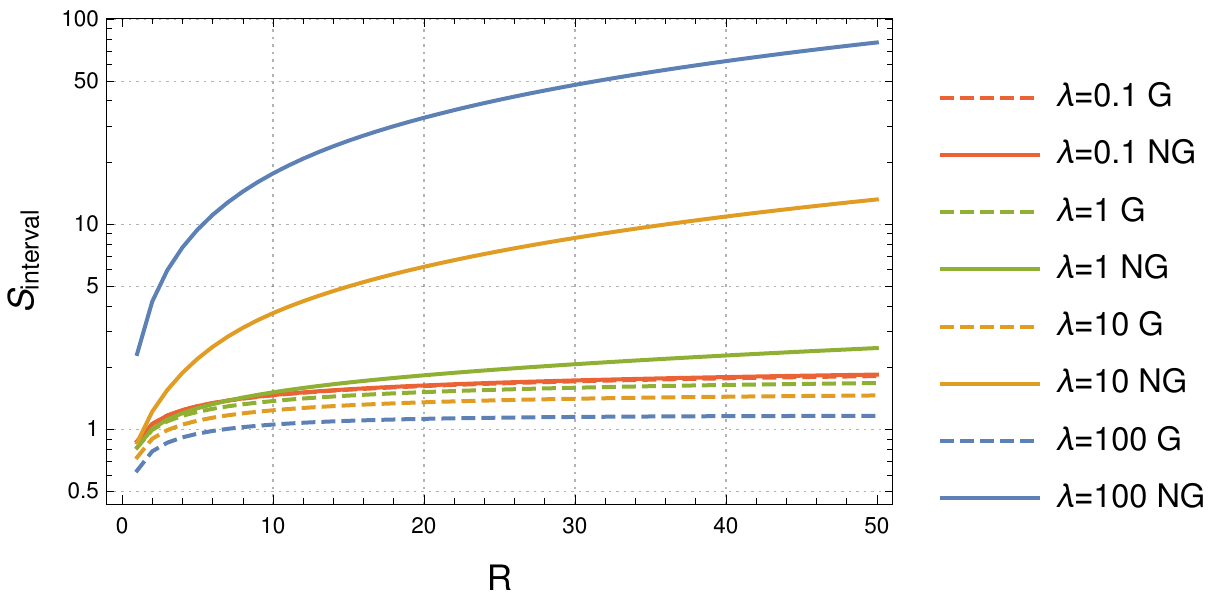}
\caption{\textit{
Entanglement entropy of intervals as a function of the size of the intervals $R$ for Gaussian (dashed) and non-Gaussian (solid) states. The latter are constructed through $m=2$ NLCT and minimize the energy functional \eqref{eq:hamiltonian} for different values of $\lambda$.
}}
\label{fig:interval-R}
\end{figure}

\begin{table}[t]
\centering
\begin{tabular}{|ccc|ccc|}
\hline
$\lambda$ & $c_2 \pm\Delta c_2$ & $\alpha\pm\Delta \alpha$ & $\lambda$ & $c_2 \pm\Delta c_2$ & $\alpha\pm\Delta \alpha$
\\
\hline
0.1 & ${-0.551\pm 0.420}$ & ${0.125 \pm 0.0417}$
&
10 & $0.294 \pm 0.013$ & ${0.941 \pm 0.009}$
\\
1 & ${0.019 \pm 0.001}$ & ${0.944 \pm 0.009}$
&
100 & ${1.742\pm 0.039}$ &  ${0.958\pm 0.005}$
\\
\hline
\end{tabular}
\caption{\textit{
Parameters $\alpha$ and $c_2$ of the nonlinear regression analysis of the entanglement entropy \eqref{eq:entr_rt_nlct} for non-Gaussian states to the nonlinear function \eqref{eq:fittings}. Different values of the coupling $\lambda$ have been considered.
}}
\label{tab:fittings}
\end{table}

In Fig. \ref{fig:interval-R} we plot $S(R)$ for various couplings $\lambda$. These curves can be studied through a nonlinear regression analysis. We assume that $S(R)$ is given by the nonlinear function
\begin{align}
S(R)= c_0 + c_1 \log R +c_2 \, R^\alpha
\ ,
\label{eq:fittings}
\end{align}
where $c_i$ and $\alpha$ are the fitting parameters. In Table \ref{tab:fittings} we show the value of $c_2$ and the exponent $\alpha$ that we obtain through nonlinear regression for different couplings. Let us note that for $\lambda\ge1$
\footnote{Let us remark that for $\lambda=0.1$ the exponent does not satisfy this pattern. It would be interesting to explore this case when other NLCT that do not break the symmetric phase ($m=3$) or other perturbative contributions are considered.}
the result $\alpha\simeq1$ and the positivity of $c_2$ give evidence of a well defined notion of mean entropy of the system \cite{Casini:2003ix}.

\begin{figure}[!t]
\centering
\includegraphics[width=\textwidth]{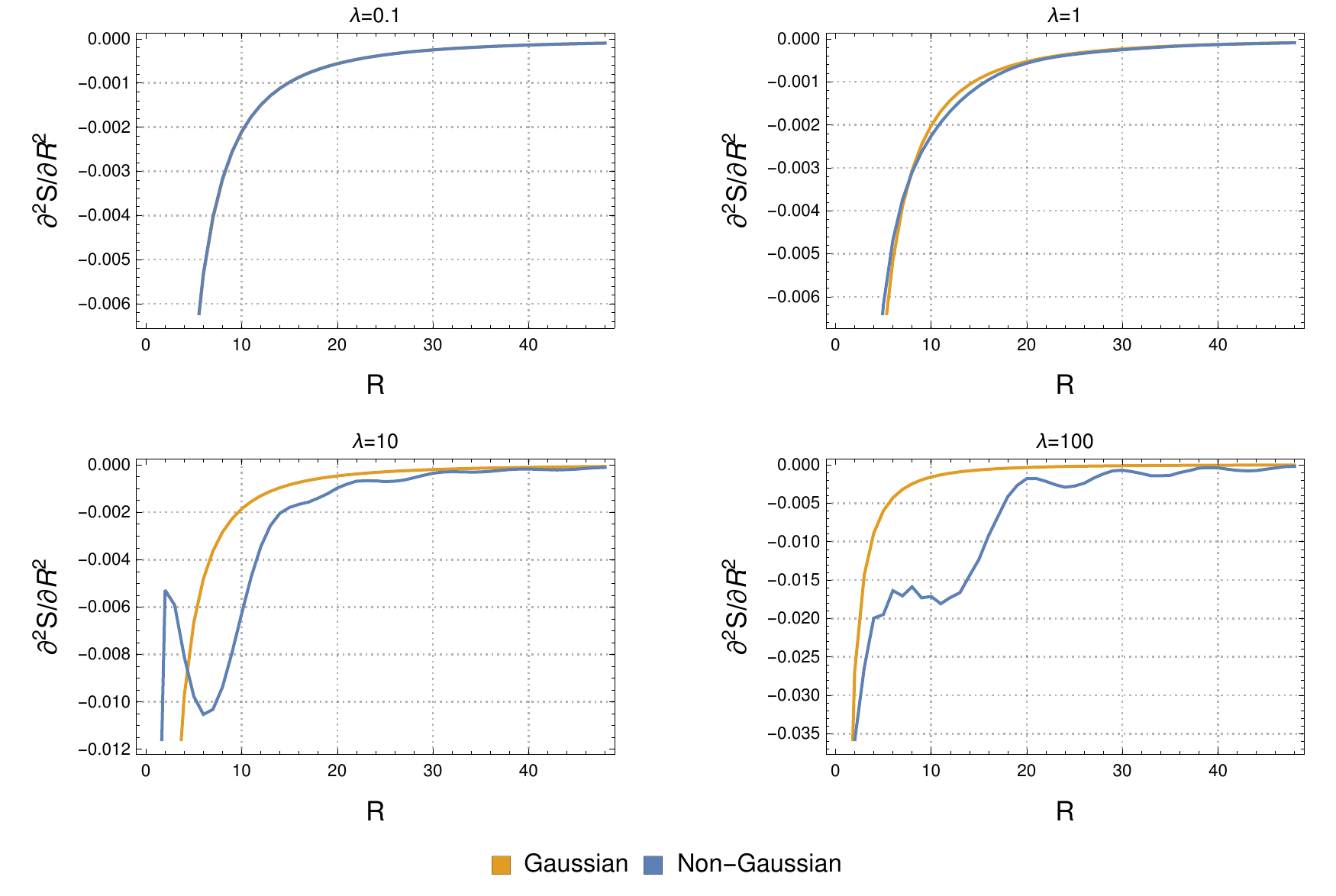}
\caption{\textit{
Concavity of the entanglement entropy for the Gaussian (orange) and non-Gaussian (blue) states that minimize the expectation value of the Hamiltonian \eqref{eq:hamiltonian} for different values of $\lambda$.
}}
\label{fig:concavity}
\end{figure}

Finally, we want to check whether the entanglement entropy for the trial non-Gaussian states satisfies some entropic properties. In particular we will study the strong subadditivity property is fulfilled. Precisely, in $1+1$ dimensions, an equivalence between strong subadditivity and concavity of the entropy can be established \cite{Casini:2003ix},
\begin{align}
\frac{\partial^2 S(R)}{\partial R^2}<0
\ .
\end{align} 
Then, when calculating this quantity we observe that the entanglement entropy clearly satisfies such condition for any various couplings that differ in various orders of magnitude. More in detail, in Figure \ref{fig:concavity} we plot the second derivative of the entropy for the interval range that we have been studied and show that it is strictly negative.

\section{Discussion and Outlook}
\label{sec:discussion}

In this work we have provided a method to study the entanglement entropy of non-Gaussian states generated by NLCT that minimize the energy functional of an interacting theory formulated at any dimension.

Our method relies on the property $\tilde\Psi[\phi]=\Psi[\Phi]$. To make use of this, we have reviewed the entanglement entropy of Gaussian states for some regions and written this quantity in terms of 2-point correlators. Therefore, we have applied such property to evaluate the entanglement entropy for non-Gaussian states. We have arrived to some formulas that are entirely general for any theory. Only the obtaining of the variational parameters through the energy minimization contains information about the interacting model that we are dealing with.

Upon considering states with NLCT \eqref{eq:BBos} for $m=2$ we have studied the entanglement entropy of the half space and intervals of size $R$ that minimize the energy functional of the $(1+1)$-dimensional $\phi^4$ theory \eqref{eq:hamiltonian}.

Let us first discuss the entanglement entropy for the half space and the dependence on the coupling. In Figure \ref{fig:half-space} our variational approximation to the entanglement entropy shows that, in the perturbative regime, it is monotonically decreasing with respect to the coupling. This result is consistent for a coupling close to zero, where the Gaussian ansatz represents an accurate approximation to the ground state of the system. Namely, in this regime, the main effect of interactions is to increase the variational mass $\mu$ through the gap equation. This implies that  field fluctuations contributing to entanglement are suppressed  much stronger than in the free case. 
\footnote{It remains to be understood whether other NLCT that are int he same phase of the perturbative expansion could shed some light on the volumetric behavior of the intervals.}
Nevertheless, one would think that this cannot be the behavior of entanglement along all the coupling regimes, thus expecting that entanglement increases as the coupling constant also grows. This is exactly the behavior shown in Figures \ref{fig:half-space} and \ref{fig:interval-R}. Namely, results indicate that there are values of the physical interaction for which the entropy turns out to be monotonically increasing with the coupling. We interpret such turning point as a value at which the higher order correlators involved in the nonlinear canonical transformation under consideration become relevant, as the entanglement entropy for Gaussian states only captures the features of the connected  2-point correlators. We strongly believe that studying this value of the coupling could be of great physical interest since it corresponds to a local minimum of entanglement entropy. This issue must be explored in more detail in subsequent works.

That said, it is important to note that for the $\phi^4$ model, it is known that for very large values of the coupling, the ground state approaches a nearly factorized mean field state with an expected monotonic decrease of the entanglement entropy towards zero. We have not explicitly explored this regime but it is important to remark that this regime is automatically accomplished by our ansatz as follows: once a cutoff $\Lambda$ is fixed, the variational parameter $g(\vp,\vq,\vr)$ in Eq.\eqref{eq:var_g} behaves differently for different values of the coupling. For the coupling regimes that we have explored, this behavior is such that as a result, the entanglement entropy, after showing the expected monotonic decrease for weak coupling, shows the monotonic increase reported above. However, from Eq \eqref{eq:var_g} and the gap equation \eqref{eq:gap_eq}, one must see that for very large values of the coupling $g(\vp,\vq,\vr) \sim \mu^{-2}$, with $\mu$ very large. In this regime, the product $s \cdot  g(\vp,\vq,\vr)\sim 0$ which makes the ansatz to automatically suppress the non-Gaussian part. This yields a Gaussian state defined by a kernel with a huge variational mass, from which it is expected a monotonic decrease of the entropy. 

When studying the entanglement entropy for intervals as a function of their size $R$, we also note its decreasing for Gaussian states as we increase $\lambda$. For the non-Gaussian states we have shown that strong subadditivity is satisfied through the concavity of the function (see Fig. \ref{fig:concavity}). On the other hand, fittings of the curves of Fig. \ref{fig:interval-R} suggest that the leading UV divergence remains proportional to $|\partial A|$. Interestingly, our results also show an extensive contribution $\propto R$.  Thus the limit $\lim_{R \to \infty} S(R)/R\geq 0$ is well defined and there exists a well defined notion of mean entropy in the system. In other words, for big enough sets the entropy is approximately extensive \cite{Casini:2003ix}. This kind of contribution typically appears when a number (scaling with volume) of IR modes  turn out to be highly entangled with the outside of the considered region and thus contribute to the entropy. By construction, an optimized nonlinear transformation acts by shifting  some of the long distance IR modes of $\phi$ by a nonlinear polynomial functional of short distance UV modes. This shifting is strongly modulated by the strength of interactions and our results indicate that the transformed IR modes are promoted to be highly entangled with the outside.

Thus, in this work we have given evidence that the entanglement entropy for non-Gaussian states satisfies some required entropic properties. Still, several questions remain to be understood. For example, we would like to study to what extent the entanglement entropy is affected depending on the parameter $m$ that determines the NLCT transformation. Additionally, it would be interesting to increase the dimensionality of the theory under consideration and check the reliability of our method through other entropic properties.

This method also allows for other further applications. For example, it is worth to mention that the approach of NLCT can be applied to fermionic field theories \cite{Fernandez-Melgarejo:2020fzw}. Despite the method can be formulated in any dimension and/or type of fermion, in  \cite{Fernandez-Melgarejo:2020fzw} we only considered the case of two-dimensional Dirac fermions. This results specially well suited to analyze the Gross-Neveu model (GN)\cite{Gross:1974jv}. In view of the results of this paper, it would be interesting to investigate the behavior of entanglement entropy in a model possessing asymptotic freedom, chiral symmetry breaking and dynamical mass generation using techniques that go beyond the Gaussian analysis \cite{Cotler:2015heb} such as those used in this work. 

Finally, in light of our results, it would also be interesting to benchmark our method and its validity by calculating other quantities for Gaussian states and checking their properties (\emph{e.g.}, complexity \cite{Jefferson:2017sdb,Bernamonti:2019zyy} or quantum relative entropy).

\section*{Acknowledgments}
The work of JJFM is supported by Universidad de Murcia-Plan Propio Postdoctoral, the Spanish Ministerio de Econom\'ia y Competitividad and CARM Fundaci\'on S\'eneca under grants FIS2015-28521 and 21257/PI/19. JMV is funded by Ministerio de Ciencia, Innovaci\'on y Universidades PGC2018-097328-B-100 and Programa de Excelencia de la Fundaci\'on S\'eneca Regi\'on de Murcia 19882/GERM/15.

\newpage


\providecommand{\href}[2]{#2}\begingroup\raggedright\endgroup

\end{document}